# A study of the solutions of the combined sine-cosine-Gordon Equation

Nan-Hong Kuo, C. D. Hu
Department of physics, National Taiwan University
Taipei, Taiwan, R.O.C.

Abstract

We have studied the solutions of the combined sine-cosine-Gordon Equation found by Wazwaz (App. Math. Comp. **177**, 755 (2006)) using the variable separated ODE method. These solutions can be transformed into a new form. We have derived the relation between the phase of the combined sine-cosine-Gordon equation and the parameter in these solutions. Its applications in physical systems are also discussed.

## 1. Introduction

The sine-Gordon equation is a very important problem[1]. It is related to many physical systems, such as, spin chains, one-dimensional superconducting arrays, and nonlinear optics. It is also a well-studied problem[2]. Although it can be solved by the inverse scattering method, various other approaches had been developed to investigate the solutions under different circumstances. Recently, a new "variable separated ODE method" was developed by Serindaoreji and Jiong[3]. It was adopted by Fu[4] and Wazwaz[5] to study the related problems. We found that, the problem studied by Wazwaz, namely the combined sine-cosine-Gordon equation is very interesting. In particular, it can be shown that the equation of motion of a spin chain with external fields can be transformed into this problem. Hence, we made further analysis and studied its application. In section 2, we briefly review the procedure of ref. 5. In section 3, we recast the solution in a physically more transparent form. The discussion of physical application is in section 4 and the conclusion is in section 5.

## 2. Analysis of solutions

In this section, we briefly review the process of ref. 5. At the same time we establish the notations. The starting point is the combined sine-Cosine-Gordon equation:

$$u_{tt} - k u_{xx} + \alpha \sin(u) + \beta \cos(u) = 0 \qquad 1$$

Here $\alpha$ and $\beta$ are arbitrary constants. Let $\xi = x - ct$ and $u = u(\xi)$, and we can change the sine-Gordon equation into an ordinary differential equation to obtain the traveling wave solution:

$$(c^2 - k) u'' + \alpha \sin(u) + \beta \cos(u) = 0 \qquad 2$$

The insight is to introduce the following form of $u(\xi)$ to satisfy above ODE

$$u'(\xi) = a \sin(\tfrac{u}{2}) + b \cos(\tfrac{u}{2}) \qquad 3$$

with the relations $ab = 2\beta/(k - c^2)$, $a^2 - b^2 = 4\alpha/(k - c^2)$. We also define $\gamma \equiv \sqrt{\alpha^2 + \beta^2} - \alpha$ for later usage.

>From ref. 5, we can obtain four kinds of solutions directly by integrating eq. (3)

$$\frac{4}{\sqrt{b^2 + a^2}} \operatorname{arctanh}\left( \frac{b \tan \tfrac{u}{4} - a}{\sqrt{b^2 + a^2}} \right) = \xi + \xi_0, \qquad 4a$$

$$\frac{4}{\sqrt{b^2 + a^2}} \operatorname{arctanh}\left( \frac{b \cot \tfrac{u}{4} + a}{\sqrt{b^2 + a^2}} \right) = \xi + \xi_0, \qquad 4b$$

$$\frac{4}{\sqrt{b^2 + a^2}} \operatorname{arctanh}\left( \frac{b \tan \tfrac{u}{4} - a}{\sqrt{b^2 + a^2}} \right) = \xi + \xi_0, \qquad 4c$$

$$\frac{4}{\sqrt{b^2+a^2}} \operatorname{arc tanh}(\frac{b\cot\frac{u}{4}+a}{\sqrt{b^2+a^2}}) = \xi + \xi_0 \qquad 4d$$

where $\xi_0$ is a constant. With the following substitution

$$\frac{\sqrt{a^2+b^2}}{4} = \frac{1}{2\sqrt{k-c^2}}\frac{\sqrt{\beta^2+\gamma^2}}{\sqrt{2\gamma}} = \frac{1}{2\sqrt{k-c^2}}\sqrt[4]{\alpha^2+\beta^2}, \qquad 5$$

we can establish the relation between eqs. (4) and the original combined sine-Gordon equation

$$u(x,t) = 4\arctan(\frac{2\sqrt[4]{\beta^2+\alpha^2}}{\sqrt{2\gamma}} \tanh[\frac{\sqrt[4]{\alpha^2+\beta^2}}{2\sqrt{k-c^2}}(x-ct)+\xi_0)] + \frac{\beta}{\gamma}) \qquad 6a$$

$$u(x,t) = 4\operatorname{arccot}(\frac{2\sqrt[4]{\beta^2+\alpha^2}}{\sqrt{2\gamma}} \tanh[\frac{\sqrt[4]{\alpha^2+\beta^2}}{2\sqrt{k-c^2}}(x-ct)+\xi_0)] - \frac{\beta}{\gamma}) \qquad 6b$$

$$u(x,t) = 4\arctan(\frac{2\sqrt[4]{\beta^2+\alpha^2}}{\sqrt{2\gamma}} \coth[\frac{\sqrt[4]{\alpha^2+\beta^2}}{2\sqrt{k-c^2}}(x-ct)+\xi_0)] + \frac{\beta}{\gamma}) \qquad 6c$$

$$u(x,t) = 4\operatorname{arccot}(\frac{2\sqrt[4]{\beta^2+\alpha^2}}{\sqrt{2\gamma}} \coth[\frac{\sqrt[4]{\alpha^2+\beta^2}}{2\sqrt{k-c^2}}(x-ct)+\xi_0)] - \frac{\beta}{\gamma}). \qquad 6d$$

At first, these four solutions looks independent. But there are actually relations between them. Equation (6b) can be transformed into eq. (6a) by $u \to u - 2\pi$, and $x \to -x$, $t \to -t$. So can eq. (6d) be transformed into eq. (6c). Hence, there are two kinds of solutions, eqs. (6a) and (6c).

## 3. Further study of the forms involving $\phi$

In this section, we will make further analysis and establish the relation between solutions in ref. 5 and the conventional solution of the sine-Gordon equation. Additionally, we will show that the solutions can be written in the form

$$u(x,t) = -\varphi + 4\arctan\frac{F(x)}{G(t)} = 4\arctan\frac{F(x)\cos\frac{\varphi}{4} - G(t)\sin\frac{\varphi}{4}}{F(x)\sin\frac{\varphi}{4} + G(t)\cos\frac{\varphi}{4}}. \qquad 7$$

We can recast solutions eqs (4) in different forms. They are more elegant and able to provide more physical insight as we will discuss later. Set $\alpha^2 + \beta^2 = R^2$ and define $\varphi$ in the following way:

$$\alpha = R\cos\varphi, \qquad 8a$$

and

$$\beta = R\sin\varphi. \qquad 8b$$

With the relation $2\sqrt[4]{\alpha^2+\beta^2}/\sqrt{2r} = 1/|\sin\frac{\varphi}{2}|$, the four solutions become

$$u(x,t) = 4\arctan\{\frac{1}{|\sin\frac{\varphi}{2}|}\tanh[\frac{\sqrt{R}}{2}(x-ct)] + \cot\frac{\phi}{2}\} \qquad 9a$$

$$u(x,t) = 4\arccot\{\frac{1}{|\sin\frac{\varphi}{2}|}\tanh[\frac{\sqrt{R}}{2}(x-ct)] - \cot\frac{\phi}{2}\} \qquad 9b$$

$$u(x,t) = 4\arctan\{\frac{1}{|\sin\frac{\varphi}{2}|}\coth[\frac{\sqrt{R}}{2}(x-ct)] + \cot\frac{\phi}{2}\} \qquad 9c$$

$$u(x,t) = 4\arccot\{\frac{1}{|\sin\frac{\varphi}{2}|}\coth[\frac{\sqrt{R}}{2}(x-ct)] - \cot\frac{\phi}{2}\}. \qquad 9d$$

$\phi$ can be viewed as an angle parameter. It has subtle physical meaning while applying the solutions to physical systems. We will discuss it based on eqs. (9) in section 4. First, we consider the following two cases:

**Case 1.** $a = 0$.

With the relation $\int du/\cos(u/2) = 4 arctanh(\tan\frac{u}{4})$, we found that the solution in eq. (4a) is reduced to

$$u(x,t) = 4\arctan[\tanh\frac{b(x-ct+\xi_0)}{4}] \qquad 10$$

which is the familiar traveling wave solution.

**Case 2.** $b = 0$.

It seems that the solutions in eqs. (4) can not be reduced to the familiar forms. However, the situation is made clearer with the following lemma:

**Lemma 1**

$$\lim_{x\to 0} arctanh(x-1) \simeq \lim_{x\to 0} \frac{1}{2}\ln(\frac{x}{2}) \qquad 11$$

**Proof** Let $y = arctanh(x-1)$ and thus $x - 1 = \tanh(y) = \frac{e^y - e^{-y}}{e^y + e^{-y}}$.

As $x \to 0$, one should have $y \to -\infty$.

Hence, $\lim_{y\to-\infty}\frac{e^y-e^{-y}}{e^y+e^{-y}} \simeq \lim_{y\to-\infty} -\frac{1-e^{2y}}{1+e^{2y}} \simeq \lim_{y\to-\infty} -(1-2e^{2y})$, and $\lim_{y\to-\infty} x \simeq \lim_{y\to-\infty} 2e^{2y}$.

As a result, $y = \frac{1}{2}\ln(\frac{x}{2})$ as $x \to 0$ and $y \to -\infty$.

**End of Proof**

With **Lemma 1**, we found that eq. (4a), in the limit $b \to 0$, becomes

$$\xi + \xi_0 = \lim_{b\to 0}\frac{4}{\sqrt{b^2+a^2}}arctanh(\frac{b\tan\frac{u}{4}-a}{\sqrt{b^2+a^2}}) = \frac{4}{a}\lim_{b\to 0}arctanh(\frac{b}{a}\tan\frac{u}{4}-1)$$
$$= \lim_{b\to 0}\frac{2}{a}\ln(\frac{b}{2a}\tan\frac{u}{4}). \qquad 12$$

To proceed, in the limit of $b \to 0$,

$$\xi + \xi_0 = \frac{2}{a}\ln\frac{b}{2a} + \frac{2}{a}\ln(\tan\frac{u}{4}). \qquad 13$$

The constants can be made canceling each other and we reached

$$u(x,t) = 4\arctan(e^{a\xi/2}). \qquad 14$$

It is just the standard traveling solution of sine-Gordon equation if $a$ is scaled away.

In next lemma, we will show that the solutions in eqs. (4) can be transformed into the expression in eq. (9).

**Lemma 2**

$$u(\xi) = 4\arctan\{\frac{1}{|\sin\frac{\varphi}{2}|}\tanh[\frac{1}{4}(\xi+\xi_0)] + \frac{\cos\frac{\varphi}{2}}{\sin\frac{\varphi}{2}}\} = -\varphi + 4\arctan(e^{(\xi-\xi'_0)/2}), \text{ if } \sin\frac{\varphi}{2} \neq 0. \qquad 15$$

**Proof** We start from eq. (4a) or (9a):

$$u(x,t) = 4\arctan\{\frac{1}{|\sin\frac{\varphi}{2}|}\tanh[\frac{\sqrt{R}}{2}(\xi+\xi_0)] + \frac{\cos\frac{\varphi}{2}}{\sin\frac{\varphi}{2}}\}.$$

For simplicity, we choose $R = 1/4$. This does not lose generality because one can always make scaling of $\xi$. With the equation

$$\tanh(\frac{\xi+\xi_0}{4}) = \frac{e^{(\xi+\xi_0)/2} - 1}{e^{(\xi+\xi_0)/2} + 1}$$

we obtain

$$\tan\frac{u}{4} = \frac{1}{\sin\frac{\varphi}{2}}\frac{e^{(\xi+\xi_0)/2} - 1}{e^{(\xi+\xi_0)/2} + 1} + \frac{\cos\frac{\varphi}{2}}{\sin\frac{\varphi}{2}} = \frac{2[e^{(\xi+\xi_0)/2}\cos^2(\frac{\varphi}{4}) - \sin^2(\frac{\varphi}{4})]}{\sin\frac{\varphi}{2}[e^{(\xi+\xi_0)/2} + 1]}$$

This can be written as

$$\tan\frac{u}{4} = \frac{\cot(\phi/4)e^{(\xi+\xi_0)/2} - \tan(\phi/4)}{e^{(\xi+\xi_0)/2} + 1} = \tan(\arctan[e^{\xi/2}(e^{\xi_0/2}\cot\frac{\phi}{4})] - \frac{\phi}{4})$$

Hence, we get

$$u(x,t) = -\phi + 4\arctan(e^{(\xi+\xi'_0)/2}) \qquad 16$$

**End of Proof**.
In equation (16),

$$\xi'_0 = \xi_0 + 2\ln(\cot\frac{\phi}{4}). \qquad 17$$

Other solutions in eqs. (4) or (9) can be transformed in a similar way.

Hence, we have shown that the solutions in eqs. (4) by Wazwaz[5] can be changed into the form of eq. (7). A more transparent form is in eq. (15) or eq. (16). One can see that the combined sine-cosine-Gordon equation has an extra phase. We have derived the relation between the phase and the parameter in eq. (4). Since $\xi_0$ can be seen as the phase or the reference point of the traveling wave solutions, or the solitons, it is interesting to observe that the presence of $\phi$ change its value. Its physical implication will be discussed in next section.

## 4. Application to spin chain systems

In this section we consider the application of the solutions in eqs. (4) or (9) to physical systems. A concrete example is an one-dimensional spin 1/2 chain[6,7]. The interaction between nearest neighbors are described by the Heisenberg model. Additionally, a controlled dimerization amplitude and applied magnetic field are also introduced. The total Hamiltonian has three parts[6]:

$$H(t) = J\sum_{i=1}^{N}\mathbf{S}_i \cdot \mathbf{S}_{i+1} + \frac{\Delta(t)}{2}\sum_{i=1}^{N}(-1)^i(S_i^+ S_{i+1}^- + S_i^- S_{i+1}^+) + h_{st}(t)\sum_i(-1)^i S_i^z. \qquad 18$$

$\Delta(t)$ is the strength of dimerization. It is a bond alternation term which can be induced by applying an electric field to the spin chain to alter the exchange interaction. $h_{st}(t)$ is the coupling of the system to a staggered external magnetic field. Since both $\Delta(t)$ and $h_{st}(t)$ come from external field, they can be varied adiabatically so as to create a parameter space. We write $\Delta$ and $h_{st}$ as $(h_{st}, \Delta) = R'(\cos\phi', \sin\phi')$, with $R'$ fixed. We shall see what effects the adiabatical variation of $\phi'$ can bring us.

Two standard methods for one dimensional systems are utilized. At first, we make the Jordon-Winger transformation to represent spins by fermion field $f_i$ and $f_i^\dagger$. Then, the bosonizations of $f_i$ and $f_i^\dagger$ will performed. As a result,

$$S_j^z = f_j^\dagger f_j - \frac{1}{2}$$
$$= \frac{\partial_z \theta_+(x_j)}{2\pi} - (-1)^j \frac{1}{\pi\alpha} \sin\theta_+(z_j) \qquad 19$$

and

$$S_i^+ S_{i+1}^- + S_i^- S_{i+1}^+ = f_j^\dagger f_{j+1} + f_{j+1}^\dagger f_j$$
$$= -\alpha[4\pi\Pi^2 + \frac{1}{4\pi}(\partial_z\theta_+)^2] - (-1)^j \frac{1}{\pi\alpha} \cos\theta_+(z_j). \qquad 20$$

Here $\theta_+$ is the bosonization phase and $\widehat{\Pi}(z) = -\partial_z\theta_+(z)$ is the conjugate momentum of $\theta_+(x)$ and $\alpha$ is the lattice constant. The Hamiltonian becomes

$$H = \int dz \{v[\pi\eta\Pi^2 + \frac{1}{4\pi\eta}(\partial_z\theta_+)^2] - \frac{R'}{\pi\alpha^2} \sin(\theta_+ + \varphi') + \frac{J}{2\pi^2\alpha^3} \cos 2\theta_+\} \qquad 21$$

where the velocity $v = (J/2)\sqrt{1 + 2/\pi\alpha}$ and the Luttinger liquid parameter $\eta = 2\sqrt{\pi\alpha/(2 + \pi\alpha)}$. The last term is irrelevant in the sense of renormalization group[6]. We thus have dropped it. The resulting equation of motion is

$$\partial_{tt}\theta_+ = v^2 \partial_{zz}\theta_+ + R\sin(\theta_+ + \varphi) \qquad 22$$

with the substitution $\phi' = \phi - \pi/2$ and $R' = \alpha^2 R/2\eta$. Apparently, space-time can be rescaled so as to get the identical sine-Gordon equation in eq. (16).

In last section we have shown that the solutions in eq. (4) or (9) can be reduced to the form in eq. (1). Note also that the position of the reference point $\xi_0$ is shifted to $\xi_0'$ in the presence of $\phi$. Physically, it means that changing the external field, or varying $\phi$, is going to shift the position of the solitons even if one considers the static case. $\theta_+$ is related to the spin operators in eqs. (19) and (20). In particular, $S^z$ contains the space derivative of $\theta_+$. Its magnitude is large at the position of solitons. One can clearly see that changing the external field $\phi$ adiabatically can move spins.

## 4. Conclusions

We have discussed the solutions of Wazwaz [5] We also have shown that they can be transformed into a more physically transparent form in eqs. (9) where a phase $\phi$ is introduced. The reference point of the soliton $\xi_0$ is affected by $\phi$. The form, when applied to a spin chain system, exhibits interesting property. The phase $\phi$ is related to external fields. By varying external fields adiabatically, and hence $\phi$, the $\xi_0$ is also changed. This implies that the solitons are moved by external fields. Since the solitons, in turn, are related the spins. We are able to reach the conclusion that the spins can be transported by external fields.